\documentclass{ws-procs9x6}

\begin{document}

\title{The Generalized Parton Distributions program at Jefferson Lab}

\author{F. Sabati\'e}

\address{C.E.A. Saclay \\
DAPNIA/SPhN \\ 
91191 Gif sur Yvette, France\\ 
E-mail: fsabatie@cea.fr}

\maketitle

\abstracts{
The Generalized Parton Distributions (GPD) have drawn a lot of interest
from the theoretical community since 1997, but also from the experimental
community and especially at Jefferson Lab. First, the results for Deeply
Virtual Compton Scattering (DVCS) at 4.2~GeV beam energy have recently been
extracted from CLAS data. The single spin asymmetry shows a remarkably
clean sine wave despite the rather low $Q^2$ achievable at this energy. Two
new dedicated DVCS experiments using 6~GeV beam will run in 2003 and 2004 in
Hall~A and Hall~B respectively. Both experiments will yield very accurate
results over a wide range of kinematics, and allow for the first time
a precise test of the factorization of the DVCS process. Assuming the Bjorken
regime is indeed reached, these experiments will allow the extraction of
linear combinations of GPD's and put strong constraints on the available
phenomenological models. Upon successful completion of both experiments, a
wider experimental program at 6~GeV can be envisioned, using for instance
Deuterium targets trying to nail down the neutron and deuteron GPD's. In
addition, resonances can be probed using $\Delta$VCS where one produces 
a resonance in the final state. In a way similar to DVCS, such a process 
depends on Transition GPD's which describe in a unique way the structure of
resonances. Finally, the 12~GeV upgrade of Jefferson Lab extends
the available kinematical range, and will allow us to perform a complete, high
precision GPD program using various reactions among which, Deeply Virtual Meson
Electroproduction (DVMP) and DVCS.}

\section{Introduction}

The understanding of the structure of the nucleon is a fundamental topic. 
Despite having been studied during the past forty years, there are still many 
questions left unanswered. An example of such is the extensive debate over the 
spin structure of the nucleon ground state. Two kinds of electromagnetic observables 
linked to the nucleon structure have been considered so far.  Electromagnetic 
form factors, first measured on the proton by Hoftstader in the 1950's, 
then more recently on the neutron. Weak form factors have been 
measured in parity violating experiments. Another approach initiated in the late 60's
studies parton distribution functions via Deep Inelastic Scattering (DIS), 
and Drell-Yan processes.
 
Recently a new theoretical framework has been proposed, namely the Generalized Parton
Distributions (GPD) \cite{ji,rad}. These provide an intimate connection between the ordinary 
parton distributions and the elastic form factors and therefore contain a wealth
of information on the quark and gluon structure of the nucleon. Moreover, not only do they
depend on the usual DIS variables, the skewdness $\xi$ (linked to $x_B$) and $Q^2$, but also
on the average parton momentum fraction in the loop $x$ and the momentum transfer between 
the initial and recoil protons $t=(p'-p)^2$, giving the GPD's much more degrees of freedom
than the regular parton distributions.

In addition, it has been
shown recently \cite{burk} that the $t$ dependence of the GPD's provide information on
the transverse position of partons inside the nucleon. One could imagine the possibility
to take ``pictures'' \cite{femto} of the nucleon with a very high definition (fixed by
the virtuality of the incoming virtual photon), which would revolutionize our understanding
of the nucleon structure and confinement in general. Such a femto-picture applied to the
deuteron case for instance would yield a much clearer understanding of where quarks are
actually located inside of the deuteron.

\section{DVCS with CLAS at 4.2~GeV}

The first accurate result on DVCS has come from Jefferson Lab with the CLAS detector in Hall~B 
\cite{clas42}.
We have measured a globally exclusive beam-spin asymmetry in the reaction
$\vec{e}p\to ep\gamma$, using a 4.25~GeV longitudinally polarized electron beam on a liquid
hydrogen target. As usual with low-energy experiments, the $ep\to ep\gamma$ process is
dominated by Bethe-Heitler (BH) where photons are emitted from the incoming or scattered electron
lines. While the interesting process is actually DVCS where the photon is emitted by the proton
in response to the electromagnetic excitation by the virtual photon, the interference between
the BH and DVCS amplitude boosts the effect of DVCS and produces a large cross-section difference
for electrons of opposite helicities. In this difference, the large BH contribution drops out
and only the helicity dependent interference term remains, parametrized as $\alpha.\sin\phi +
\beta.\sin 2\phi$, where $\phi$ is the angle between the leptonic and hadronic planes. 
Note that the $\alpha$ coefficient is linear in the leading twist GPD's.
Experimentally, the Hall~B
experiment measured the relative asymmetry $A=(d^4\sigma^+-d^4\sigma^-)/(d^4\sigma_{tot})$,
which has a more complex $\phi$ dependence than the difference in cross-sections, but is much
simpler to measure in a large acceptance spectrometer such as CLAS.

\begin{figure}[th]
\centerline{\epsfxsize=6.cm\epsfbox{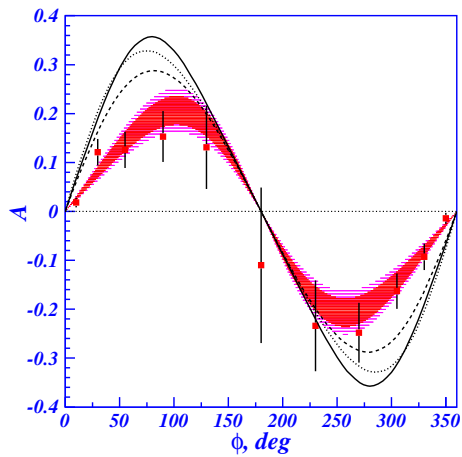}}   
\caption{$\phi$ dependence of the beam spin asymmetry A. The dark shaded region is the range of the fitted
function defined by statistical errors only. The kinematics averages to $<Q^2>=1.25$~GeV$^2$,
$<x_B>=0.19$ and $<-t>=0.19$~GeV$^2$. \label{clas42res}}
\end{figure}

The reaction $ep\to ep\gamma$ was identified by examining $ep\to ep X$ events and requiring the
mass of the missing particle to be zero. Unfortunately, CLAS is not able to separate
$\pi^0$ electroproduction from photon electroproduction using an event-by-event 
missing mass technique. The
number of photon events was determined using a fitting technique that analyzed the shape of
the missing mass distribution. The exclusivity of the reaction is therefore demonstrated globally
rather than event-by-event. Figure~\ref{clas42res} shows the resulting asymmetry $A$ as a
function of $\phi$. The data points are fitted with the function 
$A(\phi)=\alpha.\sin\phi + \beta.\sin 2\phi$ where $\alpha=0.202\pm 0.028^{stat}\pm 0.013^{syst}$ 
and $\beta=-0.024\pm 0.021^{stat}\pm 0.009^{syst}$. Up to the accuracy of the data, $\beta$ seems 
to be very small at $Q^2$ as low as 1.25~GeV$^2$,
indicating that the tranverse part of the process dominates as expected. More accurate data are
clearly needed to test the factorization in a systematic way.

\section{DVCS in Hall~A at 6~GeV}

\begin{figure}[th]
\centerline{\epsfxsize=6.cm\epsfbox{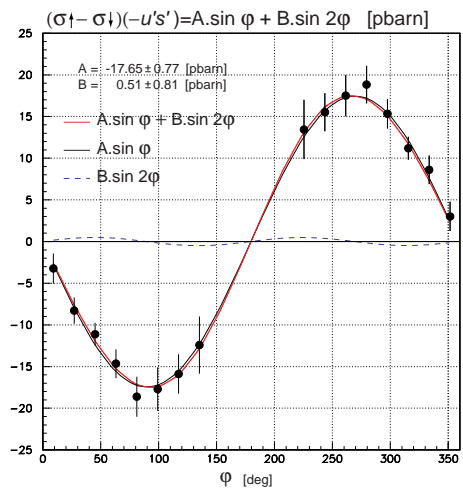}}   
\caption{Expected cross-section difference mutiplied by a kinematical factor corresponding to the BH
propagators as a function of $\phi$ for the $Q^2=2.5$~GeV$^2$ setting. Note that the convention for $\phi$
is different compared to Fig.~\ref{clas42res}. \label{reshalla}}
\end{figure}

The Hall~A DVCS experiment \cite{halla} proposes to check the factorization of the DVCS process by performing
an accurate measurement of the (properly) weighted cross-section difference for three values of $Q^2$ from 1.5 up to
2.5~GeV$^2$ at fixed $x_B=0.35$. The high resolution and high luminosity achievable in Hall~A
allows one to make a very clean interpretation of the data, since the exclusivity will be
checked on an event-by-event basis. The
experimental apparatus is composed of a High Resolution Spectrometer for the detection
of the scattered electron, a high resolution PbF$_2$ calorimeter for the detection of the emitted
photon and an array of plastic scintillators for the detection of the recoil proton. An example
of the quality of the data achievable by this experiment is shown on Figure~\ref{reshalla} for
the $Q^2=2.5$~GeV$^2$ setting.

\section{DVCS in Hall~B at 6~GeV}

\begin{figure}[th]
\centerline{\epsfxsize=10.cm\epsfbox{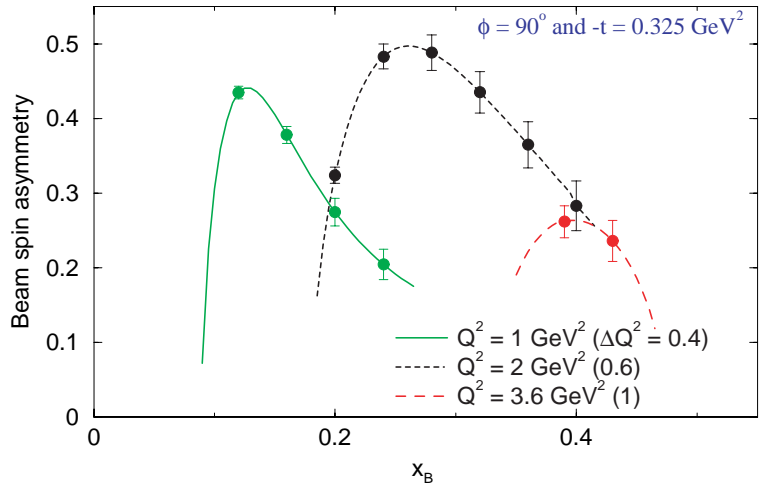}}   
\caption{$x_B$ dependence of the single spin asymmetry at $\phi=90^\circ$ expected in
the 6~GeV CLAS DVCS experiment. Three sets of points correspond to three $< Q^2 >$.
\label{reshallb}}
\end{figure}

Once the factorization for the DVCS process has been confirmed by the Hall~A experiment, DVCS
in Hall~B will allow one to look at various kinematical dependences of the beam spin asymmetry \cite{hallb}.
Indeed, CLAS with its large acceptance will allow to scan this observable in function of
$x_B$, $t$ and $Q^2$, for a total of 378 bins with good staticstics. In order to address
the issue of the full exclusivity of DVCS events, the CLAS DVCS collaboration is designing
a forward PbWO$_4$ calorimeter to detect low angle photons typical of DVCS at small $t$.
In order to achieve a higher luminosity, a Moller shield composed of a superconducting
solenoid surrounding the target is also under design. Figure~\ref{reshallb} shows
the quality of the data which is expected from this experiment. 

\section{New experiments at 6~GeV}

Although DVCS and Deeply Virtual Meson Production are not easy experiments, one could think of
the future of these measurements using 6~GeV beam. DVCS on the Deuterium is of remarkable
interest: it allows not only to measure the coherent electroproduction of photons off the deuteron,
but also, using deuteron as a quasi-free neutron target, to measure the DVCS
reaction on the neutron. This type of experiments is essential if one envisions a full flavor
decomposition of the GPD's. However, several difficulties arise. As far as the Deuteron DVCS
(D$_2$VCS) is concerned, the recoil deuteron has to escape the target, which is more
difficult than in the proton case. This will contribute to raising the minimum $-t$ achievable by the
experiment. The cross-section has been evaluated recently \cite{d2vcs}: at low $x_B$, it is
comparable to the proton cross-section. The beam spin asymmetry is also sizeable much like
in the proton case. For the neutron case, the experiment is much more difficult. Once again,
the cross-section is only 2 to 3 times smaller, but the asymmetry is much smaller than for 
the proton, of the order of a few \%.

The GPD formalism has been extended to transition reactions where the final particle is a nucleon resonance 
\cite{delta}. Just as in the neutron case, the cross-section is smaller, although not too small, and the
asymmetry is a few \%. Even though $\Delta$VCS is a very difficult experiment, the prospects for a new kind
of baryon resonance spectroscopy makes the study worth.

\section{The GPD program with 12~GeV beam at Jefferson Lab}

The extension of the planned 6~GeV GPD program up to 12~GeV is rather straightforward. Higher energy
beam allows one to open up the kinematical coverage:  $0.1\le x_B \le 0.6$ and $1\le Q^2 \le 8$~GeV$^2$. One can
therefore imagine a complete GPD program using all kinds of Hard Exclusive reactions ($ep\to ep\gamma$, 
$ep\to ep\rho$, $eD\to eD\gamma$, $ep\to en\pi^+$, ...) on different targets, polarized or not, and looking 
at various observables such as cross-sections, single spin asymmetries or even double spin asymmetries.
Both the luminosity and/or the high resolution of the Jefferson Lab equipments will make it possible to
perform accurate measurements in a reasonnable amount of time.

\section{Conclusion and outlook}

Jefferson Lab is in a unique position to make pioneering steps towards a better understanding of the structure of
the nucleon. The 4.2~GeV CLAS data has confirmed that the DVCS single spin asymmetry is indeed large and close to
a pure sine wave, encouraging us to perform two new experiments at 6~GeV in Hall~A and Hall~B. These experiments
will not only try to justify the factorization of DVCS at moderate $Q^2$, but will give the first
strong contrains to GPD phenomenological models. Additional information can be obtained
at 6~GeV using a Deuterium target and looking either at coherent deuteron DVCS, or DVCS on quasi-free
neutrons. Finally, the 12~GeV of Jefferson Lab greatly enhances the kinematical coverage
of Deeply Exclusive processes, and a complete GPD program can be performed, which undoubtedly, will shed a new light on
the understanding of the nucleon.

\end{document}